# Effects of Defects on Thermal Transport across Solid/Solid Heterogeneous Interfaces

Ershuai Yin, Wenzhu Luo, Lei Wang, Qiang Li[*]

[a] MIIT Key Laboratory of Thermal Control of Electronic Equipment, School of Energy and Power Engineering, Nanjing University of Science & Technology, Nanjing, Jiangsu 210094, China

**Abstract:** During the fabrication of heterogeneous structures inside chips, impurities and defects are inevitably introduced. However, the mechanism by which defects affect interfacial heat transport remains unclear. In this work, a microscale thermal transport model is developed by combining first-principles calculations with Monte Carlo simulations, explicitly accounting for the effects of defects. The effects of defect concentration and location on thermal transport characteristics are investigated for four heterointerfaces: Si/SiC, GaN/SiC, Si/Diamond, and GaN/Diamond. Temperature distribution, spectral thermal conductance, average phonon scattering numbers, and interfacial thermal conductance (ITC) are compared under different conditions. The results show that, for Si/SiC, Si/Diamond, and GaN/Diamond interfaces, introducing defects weakens heat transport. Higher defect concentration leads to lower ITC. Furthermore, when defects are in SiC or Diamond, which have broader phonon spectral distributions, their impact on ITC is weaker. For the GaN/SiC interface, defects in GaN reduce ITC, while defects in SiC enhance ITC. At a defect concentration of 0.05, ITC decreases by 54.1% when defects are present in GaN, but increases by 57.2% when defects are present in SiC. This behavior arises from defect-induced phonon energy redistribution near the interface. The redistribution increases the population of low-frequency phonons, which are more capable of crossing the interface, thus enhancing heat transfer. This study enriches the fundamental understanding of thermal transport across semiconductor heterointerfaces and guides the design and fabrication of high-ITC heterostructures.

**Keywords:** Heterogeneous interface; Interfacial thermal conductance; Defects; Phonon transport; Thermal management

---

[*] Corresponding author. liqiang@njust.edu.cn (Q. Li)

# 1 Introduction

With the continuous increase in chip power density and integration, thermal management has become a critical challenge restricting chip performance and reliability [1,2]. However, multiple heterogeneous interfaces often exist inside chips. Lattice mismatch and acoustic property differences between dissimilar materials cause significant interfacial thermal resistance (ITR), which limits heat dissipation [3]. For example, experiments show that in gallium nitride (GaN) chips on Diamond substrates, the temperature rise caused by ITR accounts for more than 40% of the total temperature rise [4]. In addition, predictions indicate that for GaN-based transistors, the temperature rise induced by the ITR between the GaN buffer layer and the SiC substrate layer can reach 50% [5]. Therefore, enhancing thermal transport across heterogeneous interfaces inside chips is a key focus of current thermal management research [6].

To enhance thermal transport across semiconductor heterointerfaces, several strengthening strategies have been proposed. The two typical approaches are the insertion of interlayers to improve phonon spectrum matching and the introduction of nanostructures to increase interfacial contact area. Both strategies have been demonstrated to be feasible [7,8]. For the interlayer approach, Wang et al. [9] fabricated GaN/SiC heterostructures with AlN interlayers of different thicknesses. They found that the AlN interlayer improved the overlap of phonon density of states (PDOS) and enhanced interfacial growth quality. When the AlN thickness was 104 nm, the interfacial thermal conductance (ITC) increased by 64% compared with the case without an interlayer. Our theoretical study [10] shows that inserting an appropriate SiC interlayer in Si/Diamond heterostructures bridges low-frequency phonons in Si with mid- to high-frequency phonons in Diamond, forming a phonon bridge. When the SiC interlayer thickness is 40 nm, the ITC increases by 46.6% compared with the interface without an interlayer. For the nanostructure-based approach, Cheng et al. [11] constructed ~100 nm trapezoidal nanostructures at the Si/Diamond interface. The measured ITC increased from 63.6 $Wm^{-2}K^{-1}$ to 105 $Wm^{-2}K^{-1}$, corresponding to a 65% enhancement. Lee et al. [12] tested the ITC of Al/Si interfaces with three-dimensional pillar-array structures and found that a properly designed nanostructure enhanced ITC by 88%. Our previous theoretical study [13] also reveals the mechanism of nanostructure-enhanced interfacial transport. The multiple phonon reflections inside nanostructures create sidewall thermal conduction channels, which serve as the primary enhancement mechanism. The strengthening effect is further influenced by size effects and structural morphology.

In recent years, a new interfacial engineering strategy has been proposed to strengthen interfacial heat transfer [14]. The core idea is to modify the atomic structure near the interface, such as by atomic substitution or isotope introduction, which redistributes the phonon energy. This enables better matching of phonons in the contact materials, thereby improving cross-interface phonon transport. For example, Li et al. [15] used nonequilibrium molecular dynamics (NEMD) to study the effect of substituting Ga atoms near the GaN/SiC interface with light atoms such as boron. They found that the introduction of light atoms enhanced the coupling strength of mid- and high-frequency phonons. When the substitution concentration near the interface reached 50%, the interfacial thermal conductance (ITC) increased by 50%. Lee et al. [16] also employed NEMD to investigate isotope effects on GaN/SiC interfacial transport. They found that introducing isotopes (10% $^{71}$Ga) in GaN improved the ITC by 23%, which was attributed to an increase in low-frequency phonons caused by phonon–isotope scattering. Although this strategy shows theoretical potential for enhancing ITC, the regulation of submicron-scale interfacial structures poses significant challenges for practical implementation, and its experimental feasibility has not yet been confirmed [6].

During the fabrication of heterogeneous structures inside chips, interfaces inevitably introduce impurities and defects, including point defects (vacancies, impurity atoms, isotopes) and dislocation defects[17]. Phonon–defect scattering also redistributes phonon energy and has the potential to improve phonon spectrum overlap and enhance ITC [18]. Therefore, defect engineering can be regarded as a type of interfacial engineering strategy. At present, only a few studies have discussed the impact of defects on interfacial heat transport [19]. For example, Yang et al. [20] used NEMD simulations to investigate the effect of vacancy defect concentration (3–30%) on heat transport across GaN/Diamond interfaces. They found that interfacial transport deteriorated with increasing defect concentration, and at 30% vacancy concentration, the interfacial thermal resistance increased by 67%. Namsani et al. [21] employed molecular dynamics simulations to study the effect of defect concentration on heat transport across Au/graphene interfaces. They found that higher defect concentration increased ITC, with a maximum enhancement of 26% compared with defect-free graphene. These results suggest that the effect of defect on interfacial heat transport may vary across different heterointerfaces. However, it is still not well understood why introducing defects into different materials leads to different ITC trends, or how defects modify phonon distribution and interfacial transport. The effect of defects on interfacial thermal transport remains unclear and urgently needs to be investigated.

Therefore, this work develops a microscale thermal transport model that considers defect effects by combining first-principles calculations with Monte Carlo simulations. The effects of defect concentrations and locations on thermal transport characteristics are investigated for four heterointerfaces: Si/SiC, GaN/SiC, Si/Diamond, and GaN/Diamond. Temperature distribution, spectral thermal conductance, average phonon scattering numbers, and ITC are compared. The results reveal the mechanism by which defects influence thermal transport across heterogeneous interfaces. This study enriches the fundamental understanding of interfacial transport in semiconductors and provides guidance for the fabrication of high-ITC heterostructures.

## 2 Methodology

Fig. 1 presents the schematic of the studied semiconductor heterostructure. The structure consists of two contacting materials, denoted as Material 1 and Material 2. In general, Material 1 corresponds to the chip, and Material 2 represents the substrate. In this work, Material 1 includes two widely used chip materials, Si and GaN. Material 2 includes two high-thermal-conductivity substrate materials, SiC and Diamond. The width (W) and depth (D) of the simulation domain are both set to 100 nm. Periodic boundary conditions are applied on the four side surfaces to eliminate finite-size effects perpendicular to the heat flux direction. The height (H) of the domain is set to 200 nm, equally divided between the two materials, giving 100 nm for each. A fixed temperature of 303 K is applied at the top boundary, and 297 K is applied at the bottom boundary, producing a steady downward heat flux. The choice of H = 200 nm is made after multiple evaluations to minimize the influence of phonon–boundary scattering on interfacial transport [13]. Both Material 1 and Material 2 may contain defects that affect the interfacial thermal transport. In this study, point defects are taken as the representative type, and their influence mechanisms are analyzed.

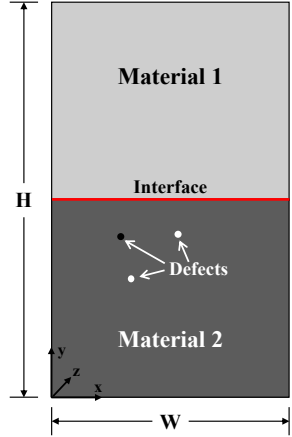

Fig. 1. Schematic of the simulation structure

The phonon transport in heterostructures is simulated by solving the deviational energy-based Boltzmann transport equation (BTE) under the relaxation time approximation (RTA) [22] using the variance-reduced Monte Carlo method [23]:

$$\frac{\partial e^d}{\partial t} + \mathbf{v}_g(k,p) \cdot \nabla e^d = \frac{\left(e^{loc} - e^{eq}_{T_{eq}}\right) - e^d}{\tau(\mathbf{k},p,T)} \tag{1}$$

where $e^d = \hbar\omega(\mathbf{k},p)\left(f - f^{eq}_{T_{eq}}\right)$ is the deviational energy distribution, $e^{loc}$ and $e^{eq}_{T_{eq}}$ are the local and equilibrium energy distributions, respectively. $\hbar$ is the reduced Planck constant, $\omega$ is the angular frequency, $\mathbf{v}_g$ is the group velocity, and $\tau$ is the relaxation time. $f$ denotes the Bose–Einstein distribution at temperature $T$, while $f^{eq}_{T_{eq}}$ is the Bose–Einstein distribution at the equilibrium temperature. $k_B$ is the Boltzmann constant.

Phonons are emitted from the isothermal boundary and undergo phonon–phonon scattering, phonon–boundary scattering, phonon–interface scattering, and phonon–defect scattering. The boundaries include isothermal and periodic types. When a phonon reaches an isothermal boundary, it disappears. When it encounters a periodic boundary, it reenters the computational domain from the opposite surface while maintaining its original properties and direction. At the interface, a phonon may be diffusely reflected or transmitted. The transmission probability at the interface is calculated as [24]:

$$\tau_{1 \to 2}(\omega') = \frac{\Delta V_2 \sum_{\mathbf{k},p} |\mathbf{v}_{g,2} \cdot \mathbf{n}| \delta(\omega' - \omega)}{\Delta V_1 \sum_{\mathbf{k},p} |\mathbf{v}_{g,1} \cdot \mathbf{n}| \delta(\omega' - \omega) + \Delta V_2 \sum_{\mathbf{k},p} |\mathbf{v}_{g,2} \cdot \mathbf{n}| \delta(\omega' - \omega)} \quad (2)$$

where subscripts 1 and 2 denote the two contacting materials. $\Delta V$ represents the volumes of the discretized cells corresponding to the Brillouin zones, and $\delta$ is the Dirac delta function.

To characterize the effect of defects on phonon transport in heterostructures, a separate phonon–defect scattering scheme is introduced into the MC simulation. Unlike phonon–phonon scattering, phonon–defect scattering is a two-phonon process, in which the phonon changes only its direction while keeping its frequency and mode unchanged. The type of scattering that occurs during phonon transport is determined based on the calculated time-to-scattering [23]:

$$\Delta t = -\tau(\omega, p, T_{eq}) \ln(R) \quad (3)$$

where $R$ is a random number with a value between 0 and 1. During the calculation, the time-to-scattering for two-phonon ($\Delta t_2$) and three-phonon ($\Delta t_3$) processes is evaluated separately. The smaller value determines the next scattering type.

In the calculation of $\Delta t_3$, the phonon relaxation time of the bulk material is used. In contrast, $\Delta t_2$ is determined by the defect-scattering relaxation time. Defects that may exist inside the material or at the interface include point defects (vacancies, impurity atoms, isotopes) and dislocation defects. Their effects can usually be incorporated into the defect-scattering relaxation time ($\tau_s$) of phonons through Matthiessen's rule. Thus, the scattering rate of defect scattering is expressed as:

$$\tau_s^{-1} = \sum_i \tau_i^{-1} \quad (4)$$

For point defects, the scattering rate is calculated as [25,26]:

$$\tau_m^{-1} = \frac{V_0 \omega^4}{4\pi v_g^3} \sum_i f_i \left(\frac{\Delta M_i}{M}\right)^2 \quad (5)$$

where $V_0$ is the average atomic volume, $f_i$ is the relative concentration of the i-th defect, and $M$ is the average atomic mass without defects. $\Delta M_i$ is the mass difference between the i-th point defect atom and the average atomic mass $M$. For vacancy defects, $\Delta M_i$ equals M. Since the scattering rates of different point defects are calculated in the same manner, the influence trends of vacancies and impurity defects are similar, with only numerical differences. Therefore, in the following analysis, vacancy defects are primarily considered as representative cases to study their influence on interfacial phonon transport.

It is worth noting that the relative defect concentration here refers to the ratio of the number of vacancy defects to the total number of atoms in the bulk material. In this study, the relative defect concentration ranges from $1\times10^{-8}$ to 0.05. Since the number of atoms contained in a unit volume differs among materials, the absolute defect concentration varies even when the relative defect concentration is the same. For Si, the vacancy defect concentration ranges from $5\times10^{14}$ cm$^{-3}$ to $2.5\times10^{21}$ cm$^{-3}$. For SiC, it ranges from $9.7\times10^{14}$ cm$^{-3}$ to $4.9\times10^{21}$ cm$^{-3}$. For GaN, it ranges from $8.8\times10^{14}$ cm$^{-3}$ to $4.4\times10^{21}$ cm$^{-3}$. For Diamond, it ranges from $1.8\times10^{15}$ cm$^{-3}$ to $8.8\times10^{21}$ cm$^{-3}$.

To ensure computational accuracy, all phonon parameters in this study are obtained from first-principles calculations. We first compute the second- and third-order force constants of the materials by combining the Vienna Ab initio Simulation Package (VASP) [27,28], Phonopy [29], and Thirdorder packages [30]. These force constants are then used in the almaBTE package [31] to calculate all required phonon properties. In the first-principles calculations, the LDA functional is adopted, the plane-wave cutoff energy is set to 520 eV, and the precision is set to High. For Si, Diamond, SiC, and GaN, the supercell sizes are 5×5×5, 6×6×6, 4×4×4, and 5×5×5, respectively. The maximum interaction range for third-order force constant calculations is the sixth nearest neighbor. In almaBTE calculations of phonon properties, a 15×15×15 phonon **q**-mesh is used for all materials.

In the MC simulations, the total number of phonons is set to $6\times10^5$, and the equilibrium temperature is 300 K. The grid size in the x, y, and z directions is 20×100×20 [10]. After the MC calculation is completed, the temperature and heat flux distributions of the heterostructure are obtained. The interfacial thermal conductance (ITC) is then calculated using the following expression [32,33]:

$$G = \frac{Q}{\Delta T} \quad (6)$$

where $Q$ is the heat flux across the interface, and $\Delta T$ is the average temperature drop at the interface.

In addition to interfacial transport properties, the total thermal resistance of the heterostructure is also calculated to evaluate the effect of defects on overall thermal performance:

$$R = \frac{\Delta T_s}{Q} \quad (7)$$

where $\Delta T_s$ is the total temperature difference of the heterostructure, which is set to 6 K in all calculations.

To further analyze the influence of defects on phonon scattering, two statistical parameters are defined: the average number of three-phonon scattering events and the average number of defect-scattering events, given as:

$$N_{ave,3} = \frac{\sum_{Phonons} N_{three-phonon\ scattering}}{N_{Phonons}} \tag{8}$$

$$N_{ave,2} = \frac{\sum_{Phonons} N_{defect-phonon\ scattering}}{N_{Phonons}} \tag{9}$$

where $N_{Phonons}$ is the total number of phonons. $N_{three-phonon\ scattering}$ denotes the number of three-phonon scattering events experienced by a phonon from emission to disappearance. $N_{defect-phonon\ scattering}$ denotes the number of defect-scattering events experienced by a phonon, which equals zero in the absence of defects.

## 3 Results and Discussion

### 3.1 Model verification

Fig. 2(a) first compares the calculated thermal conductivities of Si, GaN, SiC, and Diamond at different temperatures with experimental data. The calculated results for all four materials agree well with experiments, confirming the accuracy of the phonon properties obtained from first-principles calculations. Subsequently, the ITCs of the four heterointerfaces are calculated using our MC method and compared with the best available experimental results and reported NEMD results, as shown in Fig. 2(b). For GaN/Diamond and Si/Diamond interfaces, our results are very close to experimental values and show higher accuracy than NEMD data. For the Si/SiC interface, the calculated ITC is also closer to experimental data compared with NEMD. For the GaN/SiC interface, our results are consistent with reported NEMD values, and both exceed the experimental data, suggesting significant room for experimental process optimization. Overall, these results demonstrate that our model predicts ITC of heterostructures with high accuracy, in close agreement with experiments.

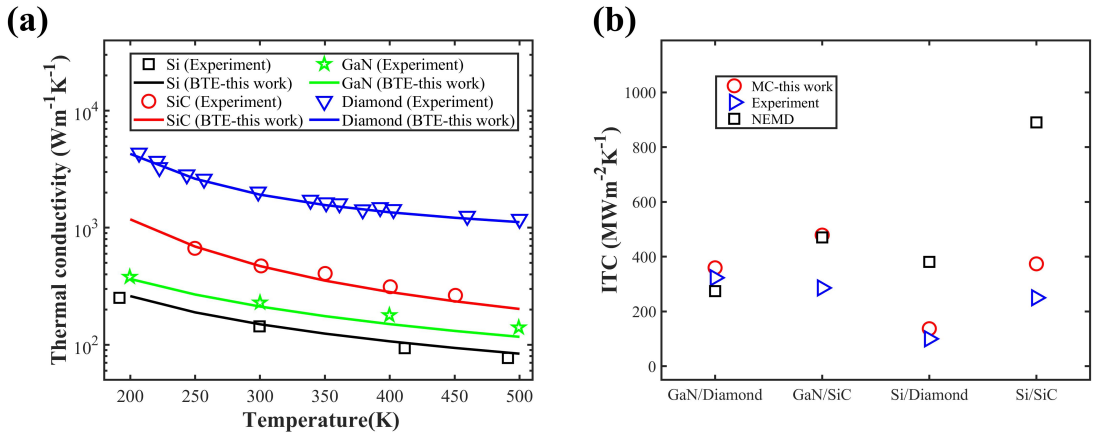

Fig. 2. Model validation. (a) Comparison of the calculated and measured thermal conductivities of Si, GaN, SiC, and Diamond. Experimental data are from Ref. [34] for Diamond, Ref. [35] for Si, Ref. [36] for SiC, and Ref. [37] for GaN. (b) Comparison of ITCs for four heterointerfaces obtained by different methods. Calculated and experimental results are from Refs. [7,38].

*3.2 Effect of Defects on interfacial thermal transport in Si/SiC, Si/Diamond, and GaN/Diamond heterostructures*

We first take the Si/SiC heterointerface as an example to illustrate the mechanism by which defects influence interfacial heat transport. Fig. 3(a) shows the temperature distributions for three cases: a defect-free structure, a structure with vacancy defects of 0.01 concentration in Si, and one with defects in SiC. In all three cases, a significant temperature drop appears at the interface, with values of 2.44 K, 1.98 K, and 2.03 K, respectively. Since the total temperature difference is fixed at 6 K in the simulations, the ratios of interfacial temperature drop to total temperature difference are 40.7%, 33.0%, and 33.8%. For defective cases, the interfacial temperature drops ratio decreases. This is because defect scattering weakens phonon transport inside the bulk materials, thereby increasing their thermal resistance. This trend is also evident from the steeper temperature gradients within the defective bulk regions. Fig. 3(b) compares the ITC and total thermal resistance of the heterostructures under the three conditions. Compared with the defect-free case, introducing defects significantly reduces interfacial thermal transport performance. When Si contains defects at a concentration of 0.01, the ITC decreases by 22.2%. When SiC contains the same concentration of defects, the ITC decreases by 8.7%. In addition, defects increase the total thermal resistance of the heterostructure. Compared with the defect-free case, the total resistance increases by 58.3% when defects are in Si, and by 31.7% when defects are in SiC.

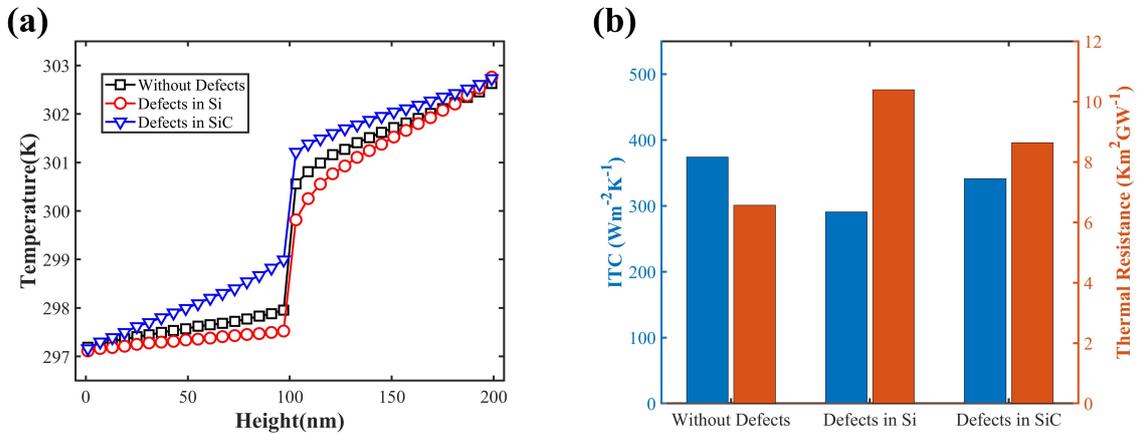

Fig. 3. Comparison of thermal characteristics between defect-free and defective Si/SiC heterostructure with a defect concentration of 0.01. (a) Temperature distributions. (b) ITC and total thermal resistance.

Fig. 4 presents the three-phonon scattering relaxation times and the defect-scattering relaxation times at a concentration of 0.01 for Si and SiC. The general trend is that low-frequency acoustic phonons usually have longer relaxation times, whereas high-frequency phonons have shorter relaxation times. Since low-frequency phonons typically possess higher group velocities, they exhibit longer mean free paths and usually dominate heat transport inside bulk materials. The frequency dependence of defect-scattering relaxation times shows a similar trend to that of three-phonon scattering, decreasing with increasing frequency. This indicates that high-frequency phonons are more likely to undergo defect scattering compared with low-frequency phonons, thereby impeding heat transfer. As a result, the presence of defects primarily suppresses the contribution of high-frequency phonons to thermal transport.

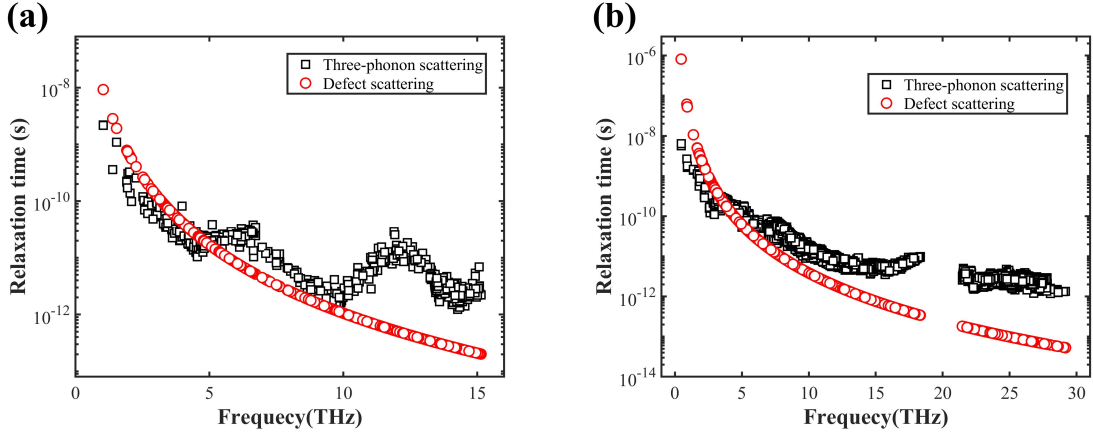

Fig. 4. (a) Si and (b) SiC relaxation times for three-phonon and defect scattering. The three-phonon relaxation times are obtained from first-principles calculations, while the defect-scattering relaxation times are calculated using Eq. (5). Results are shown for a defect concentration of 0.01.

Fig. 5(a) shows the phonon density of states (PDOS) of bulk Si and SiC, calculated from first principles. A strong overlap of PDOS is observed in the wide frequency range of 0–15 THz, indicating that phonons within this range can elastically transmit across the interface. However, due to the lack of corresponding phonons in Si, the SiC acoustic phonons above 15 THz and all optical phonons cannot transmit across the interface. Fig. 5(b–d) compare the spectral thermal conductance (SHC) distributions of heterostructures with and without defects. As shown in Fig. 5(b), for the defect-free case, heat transport in Si mainly originates from two frequency ranges near 5 THz and 12 THz. The first peak corresponds to acoustic phonons in Si, while the second peak corresponds to optical phonons in Si. For SiC, heat conduction is dominated by acoustic phonons in a range of 0~18 THz, while optical phonons contribute negligibly due to their low group velocity and short mean free path. Owing to phonon–interface scattering, the SHC in the interfacial region differs significantly from that in the bulk. Three distinct peaks appear in the interfacial SHC, at 5 THz, 9.5 THz, and 12 THz.

By comparing Fig. 5(b) and 5(c), when defects are in Si, extensive phonon–defect scattering drastically reduces the transport contribution of optical phonons, while slightly enhancing acoustic phonon transport. In the interfacial region, the SHC between 7~15 THz decreases sharply, with stronger suppression at higher frequencies. In contrast, transport in the 0~7 THz range is only slightly affected by defects. Similarly, comparing Fig. 5(b) and 5(d), introducing defects in SiC also suppresses high-frequency phonon transport in the bulk and reduces SHC in the 7–15 THz range at the interface. However, phonon–defect scattering strongly suppresses high-frequency phonon transport, forcing more heat to be carried by low-frequency phonons. As a result, the fraction of low-frequency phonons transmitted to the interface from SiC increases, which improves their overlap with low-frequency phonons in Si. This phonon energy redistribution explains why, at the same defect concentration, the ITC reduction is smaller when defects are introduced in SiC compared with defects in Si.

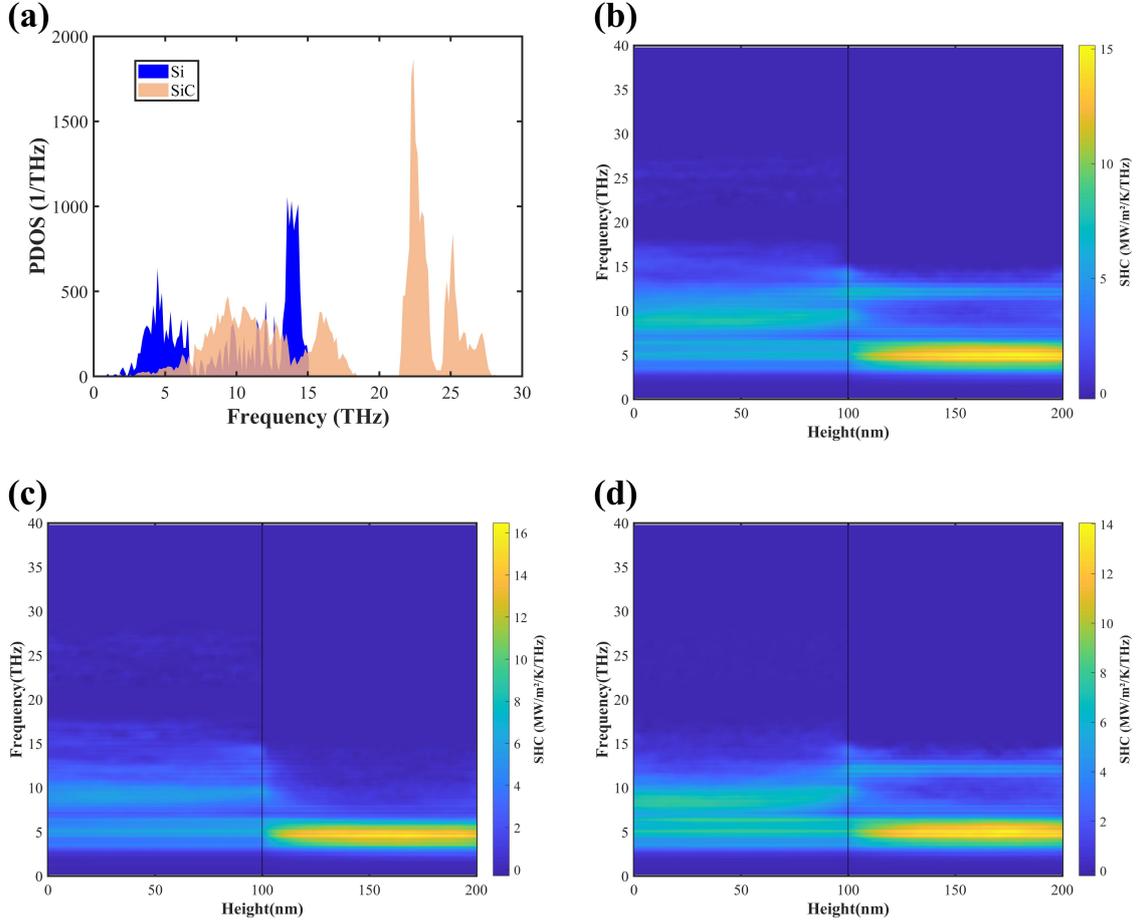

Fig. 5. (a) Comparison of the PDOS of Si and SiC. (b) SHC distribution of the defect-free Si/SiC heterostructure. (c) SHC distribution of the Si/SiC heterostructure with defects at a concentration of 0.01 in Si. (d) SHC distribution of the Si/SiC heterostructure with defects at a concentration of 0.01 in SiC.

Fig. 6 shows the effect of defect concentration on the ITC of the Si/SiC heterointerface. The results indicate that at very low defect concentrations, the ITC remains nearly identical to the defect-free case, regardless of whether defects are in Si or SiC. In some cases, the ITC is slightly higher than that of the defect-free structure, because the phonon-defect scattering can improve the contribution of low-frequency phonons. However, this enhancement is negligible. For example, when SiC contains defects at a concentration of $1 \times 10^{-8}$, the ITC increases by only 1.35% compared with the defect-free case. Therefore, introducing defects is not a feasible strategy for enhancing ITC of the Si/SiC heterointerface. As defect concentration increases, the ITC decreases rapidly, while the rate of reduction gradually slows. The influence of defects also differs depending on the location. Defects in SiC lead to a smaller ITC reduction compared with defects in Si. For instance, at a defect concentration of 0.05, the ITC decreases by 20.65% when defects are in SiC, but by 40.49% when defects are in Si. These results highlight that reducing defect concentration is essential for ensuring high ITC in Si/SiC heterointerfaces. Since defects in SiC weaken ITC to a lesser extent, growing SiC on Si may be a more favorable approach in the fabrication of Si/SiC heterostructures.

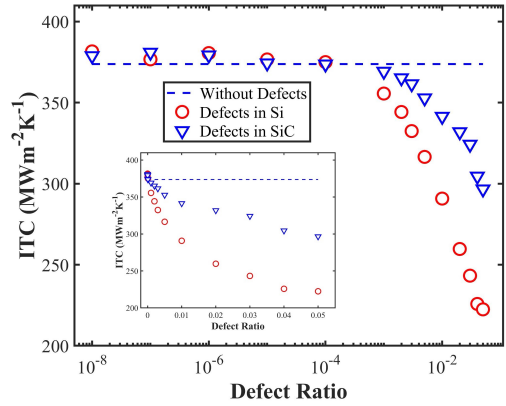

Fig. 6. Effect of defect concentration on ITC of the Si/SiC heterointerface.

To further analyze the influence of defect concentration on thermal transport, the average numbers of phonon–phonon and phonon–defect scattering events during the lifetime of phonons are calculated. As shown in Fig. 7, adding defects has little effect on the three-phonon scattering process inside the heterostructure. However, the average number of phonon–defect scattering events increases linearly with defect concentration. At the same defect concentration, phonon–defect scattering events are significantly more frequent when defects are in SiC than in Si. This is because SiC exhibits much shorter defect-scattering relaxation times in the high-frequency range (considerably lower than those of three-phonon scattering), making phonon–defect scattering events more likely. These scatterings change the phonon transport direction without altering frequency or mode. As a result, phonon–defect scattering markedly increases the thermal resistance inside the bulk materials, as also illustrated in Fig. 3(b).

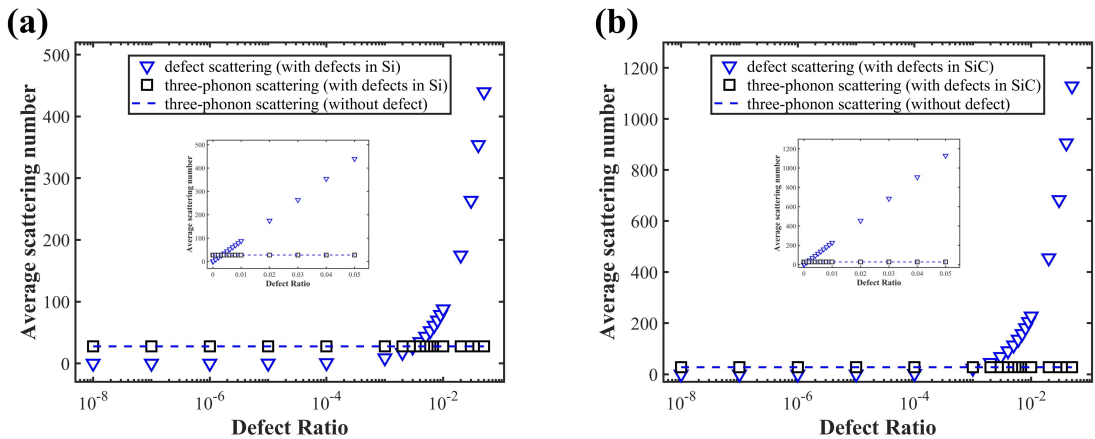

Fig. 7. Effect of defect concentration on the average number of phonon scattering events. (a) Defects in Si. (b)

Defects in SiC.

Fig. 8 shows the effect of defect concentration on the ITC of Si/Diamond and GaN/Diamond heterointerfaces. As defect concentration increases, the ITC decreases in both systems. For the Si/Diamond interface, defects in Si cause a stronger reduction in ITC. At a concentration of 0.05, the ITC decreases by 45.5% when defects are in Si, but by only 22.4% when defects are in Diamond. For the GaN/Diamond interface, defects in GaN lead to a stronger reduction. At a concentration of 0.05, the ITC decreases by 60.4% when defects are in GaN, compared with only 9.9% when defects are in Diamond. When defects are in Diamond, phonon–defect scattering suppresses high-frequency phonon transport at the interface, while increasing the number of low-frequency phonons. This shifts the phonon distribution toward lower frequencies, improving the overlap with low-frequency phonons in Si and GaN. As a result, defects in Diamond exert a weaker influence on ITC. Considering that defects are unavoidable during fabrication, growing Diamond on single-crystal Si or GaN is a feasible approach to achieve heterostructures with high ITC.

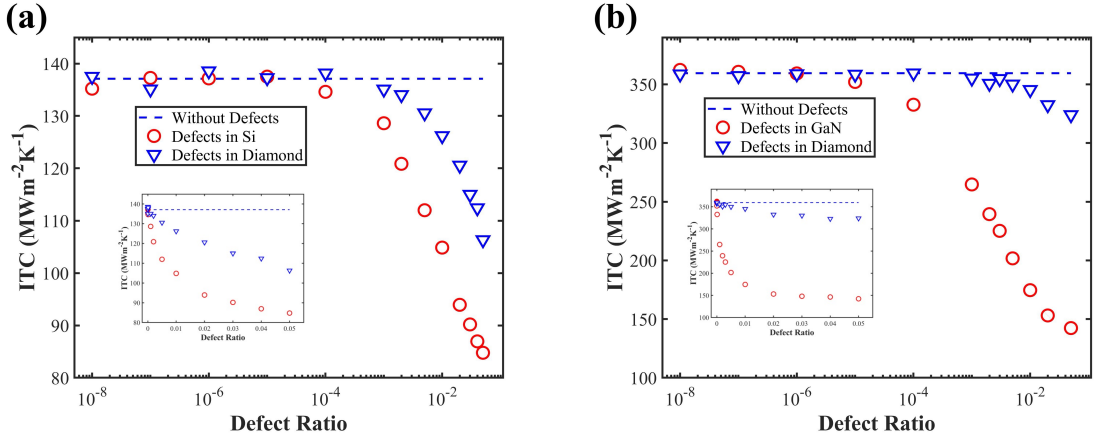

Fig. 8. Effect of defect concentration on ITC of the (a) Si/Diamond (b) GaN/Diamond heterointerface.

*3.3 Defect-induced enhancement of interfacial thermal transport in GaN/SiC heterostructures*

Fig. 9 shows the effect of defect concentrations on the thermal transport performance of GaN/SiC heterostructures. At low defect concentrations (below $1\times10^{-4}$), both the ITC and the total thermal resistance of the heterostructure remain nearly identical to those of the defect-free case, regardless of whether defects are in GaN or SiC. However, with increasing defect concentration, opposite trends appear. When defects are in GaN, the ITC decreases gradually, with the reduction trend slowing down at higher concentrations. In contrast, when defects are in SiC, the ITC increases almost linearly with defect concentration. We do not calculate results at higher defect concentrations, since excessive defects would significantly increase bulk thermal resistance and may compromise structural integrity, making the results unrealistic. At a defect concentration of 0.05, the ITC decreases by 54.1% when defects are in GaN, whereas it increases by 57.2% when defects are in SiC. Despite these opposite trends in ITC, the total thermal resistance of the heterostructure increases with defect concentration in both cases. At a defect concentration of 0.05, the total resistance increases by 245.3% when defects are in GaN and by 89.7% when defects are in SiC. The effect is stronger for GaN because defects increase both the bulk thermal resistance of GaN and the interfacial thermal resistance simultaneously.

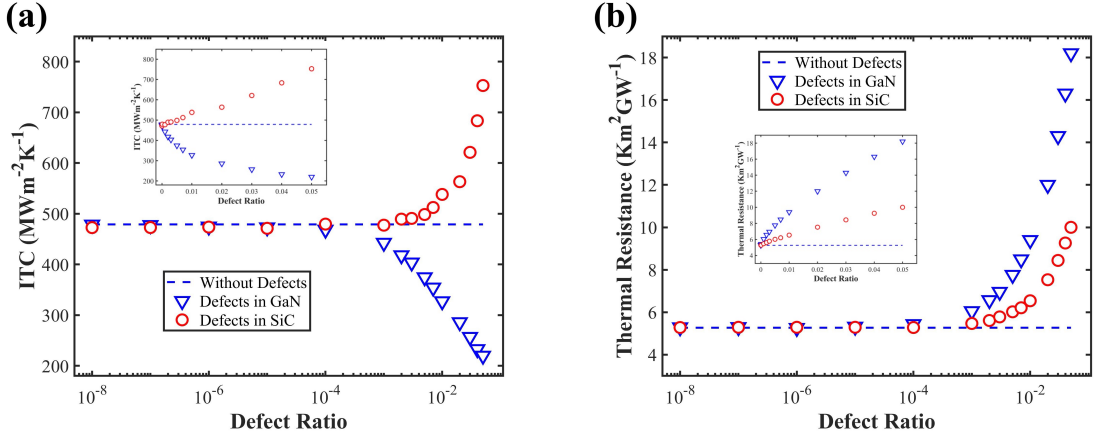

Fig. 9. Effect of defect concentration on the thermal transport performance of GaN/SiC heterostructures: (a) ITC; (b) total thermal resistance.

The enhancement of ITC by introducing defects in SiC arises from the modification of the phonon energy distribution. As shown in Fig. 10(a), GaN and SiC exhibit great PDOS overlap in the 0~10 THz range. In addition, some acoustic phonons in SiC (15~18 THz) and optical phonons near 21 THz also overlap with GaN optical phonons, suggesting that phonons in these ranges can transmit across the interface. According to Fig. 10(b), heat transport in bulk GaN mainly depends on phonons in the ranges of 0~10 THz and 15~23 THz. Owing to the higher group velocity and longer mean free path, acoustic phonons below 10 THz contribute far more heat transport than optical phonons. Because GaN lacks phonons in the 10~15 THz range, the interfacial SHC of SiC decreases in this interval. Optical phonons in SiC contribute negligibly to heat transport in both the bulk and interfacial regions. Comparing Fig. 10(b) and 10(c), when GaN contains a high concentration of defects, high-frequency phonon transport is almost completely suppressed, leaving only a single SHC peak near 5 THz. This suppression eliminates high-frequency phonon transmission in GaN, which mismatches with the broad phonon distribution of SiC, significantly reducing interfacial transport in the 5~10 THz and 15–23 THz ranges. As a result, the ITC is severely weakened. Comparing Fig. 10(b) and 10(d), introducing defects in SiC also reduces high-frequency phonon transport. However, because the mean free path of optical phonons in SiC is intrinsically very short and their contribution to heat transport is poor, the weakening effect is limited. As shown in Fig. 11(a), transmissible phonons across the GaN/SiC interface are mainly distributed in the low-frequency region below 10 THz. Defects in SiC increase the fraction of low-frequency phonons in the interfacial region, thereby raising the proportion of phonons capable of transmission across the interface and enhancing interfacial heat transport. Fig. 11(b) shows that, although the SHC of high-frequency phonons in SiC decreases slightly after defect introduction, the substantial increase in low-frequency SHC compensates for this loss. Consequently, defects in SiC strengthen the ITC of the GaN/SiC interface. However, it is worth noting that defects also significantly reduce the intrinsic thermal conductivity of SiC. Thus, controlling defect concentration within a limited range is essential to reduce the total thermal resistance of heterostructures.

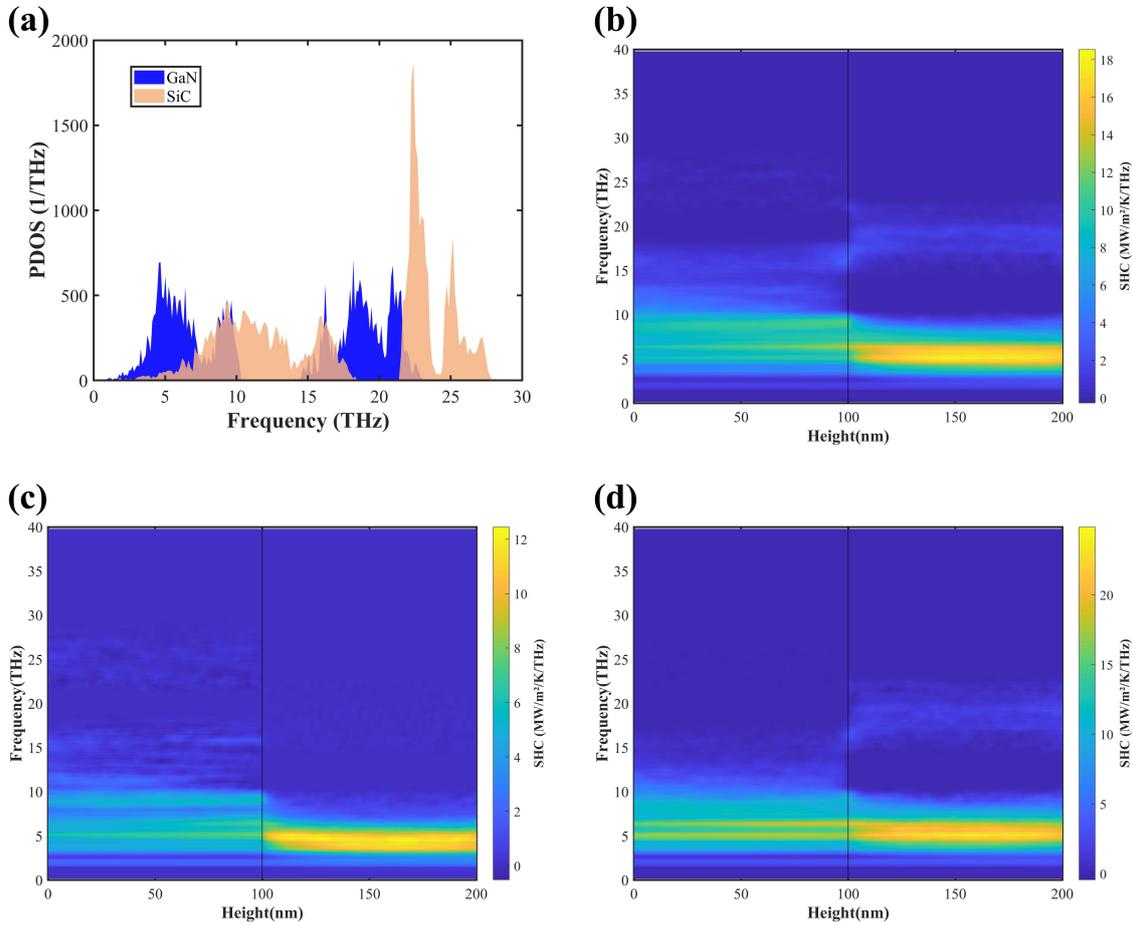

Fig. 10. (a) Comparison of the PDOS of GaN and SiC. (b) SHC distribution of the defect-free GaN/SiC heterostructure. (c) SHC distribution of GaN/SiC heterostructure with defects at a concentration of 0.03 in GaN. (d) SHC distribution of GaN/SiC heterostructure with defects at a concentration of 0.03 in SiC.

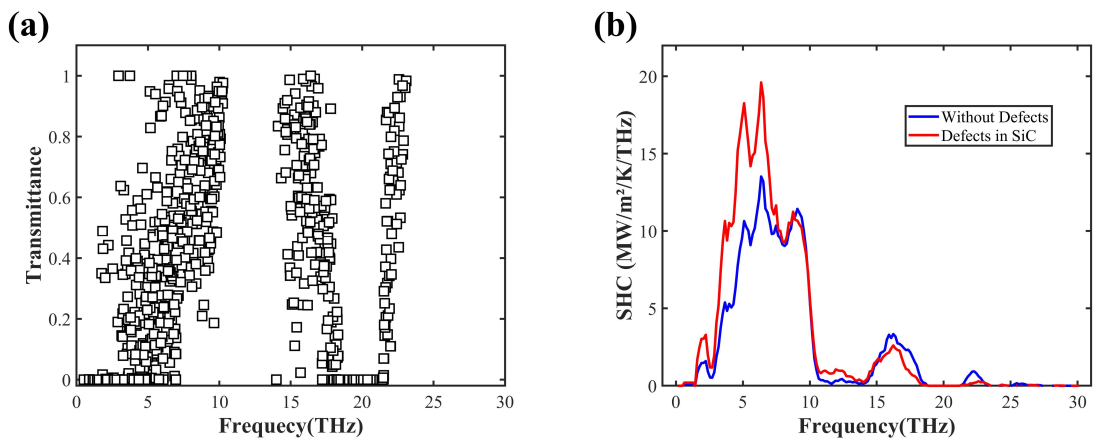

Fig. 11. (a) Phonon transmittance of the GaN/SiC interface. (b) Comparison of the SHC distributions between the defect-free GaN/SiC heterostructure and the one with defects at a concentration of 0.03 in SiC.

## 4 Conclusions

This work investigates the effects of defect concentration and location on the interfacial thermal transport of four heterointerfaces (Si/SiC, GaN/SiC, Si/Diamond, and GaN/Diamond) to reveal the mechanisms by which defects influence solid/solid interfacial heat transfer. The main conclusions are as follows:

1) Introducing defects weakens heat transport across the Si/SiC, Si/Diamond, and GaN/Diamond heterointerfaces. With increasing defect concentration, ITC decreases, and the reduction trend gradually slows. Because both bulk thermal conductivity and ITC are simultaneously degraded, the total thermal resistance of the heterostructure increases with defect concentration.

2) Owing to shorter relaxation times at high frequencies, phonon–defect scattering occurs more easily and suppresses heat transfer carried by high-frequency phonons. At the same time, the fraction of low-frequency phonons reaching the interface increases, which can enhance low-frequency phonon transmission across the interface.

3) For Si/SiC, Si/Diamond, and GaN/Diamond heterointerfaces, defects in materials with broader phonon spectra, such as SiC and Diamond, have weaker effects on ITC. For example, at a defect concentration of 0.05 in the Si/SiC interface, the ITC decreases by 20.65% when defects are in SiC, compared with a 40.49% decrease when defects are in Si. Therefore, growing SiC or Diamond on Si and GaN is a feasible strategy to obtain heterostructures with high ITC.

4) For the GaN/SiC heterointerface, defects in GaN reduce ITC, while defects in SiC enhance it. At a defect concentration of 0.05, ITC decreases by 54.1% when defects are in GaN, whereas it increases by 57.2% when defects are in SiC. This enhancement is mainly due to an increase in low-frequency phonons in the interfacial region, which raises the fraction of phonons that can transmit across the interface, thereby improving interfacial heat transport.

In summary, this study provides important insights into the mechanisms of defect-induced interfacial thermal transport and offers guidance for the fabrication of high-ITC heterostructures.

**Acknowledgments**

This work was supported by the National Natural Science Foundation of China (Grant NO. 52006102), the Fundamental Research Funds for the Central Universities (No.30923010917).